\input{aipcheck}

\documentclass[
    ,final            
  ]
  {aipproc}

\layoutstyle{6x9}

\begin{document}

\title{Stochastic Time}


\author{Toru Ohira}{
  address={Sony Computer Science Laboratories, Inc., Tokyo, Japan 
141-0022}
}

\begin{abstract}
 We present a simple dynamical model to 
address the question of introducing a stochastic nature in a time variable. 
This model includes noise in the time variable but not in the ``space" 
variable, which is opposite to the normal description of stochastic dynamics. 
The notable feature is that these models can induce a ``resonance" with 
varying noise strength in the time variable. Thus, they provide a different 
mechanism for stochastic resonance, which has been discussed within the 
normal context of stochastic dynamics.
\end{abstract}

\maketitle


``Time" is a concept that has gained a lot of attention from thinkers in 
virtually all disciplines\cite{davies1995}. In particular,  
our ordinary perception of time is not the same as that of space, and this difference has been 
appearing in a variety of contemplations about nature. It appears to be the main 
reason for the theory of relativity, which has conceptually brought space and 
time closer to receiving equal treatment, continues to fascinate and attract 
discussion in diverse fields. Also, issues such as ``directions" or 
''arrows" of time are current interests of research\cite{savitt1995}. Another 
manifestation of this difference is the treatment of noise or fluctuations in dealing 
with dynamical systems. When we consider dynamical systems, whether classical, quantum, or 
relativistic, time is commonly viewed as not having stochastic characteristics.
In stochastic dynamical theories, we associate noise 
and fluctuations with only ``space'' variables, such as the position of a 
particle, but not with the time variables. In quantum mechanics, the concept of 
time fluctuation is well accepted through the time-energy uncertainty 
principle. However, time is not treated as a dynamical quantum observable, 
and a clearer understanding has been explored\cite{Busch2002}.  

Against this background, our main theme of this paper is to consider 
fluctuations of time in classical dynamics through the presentation of a simple 
model. There are variety of ways to bringing stochasticity to some temporal aspects of
 dynamical systems. The model which we present is one way, it is an extension of delayed dynamical models\cite{mackey77,cooke82,milton89,ohira-yamane00,frank-beek01}.
With stochastic time, we have found that the model exhibits behaviors similar to those
investigated in the topic of stochastic 
resonance\cite{wiesenfeld-moss95,bulsara96,gammaitoni98}, which are studied in a variety of 
fields\cite{mcnamara88,longtin-moss91,collins1995,chapeau2003,lee2003}. The difference is
that the phenomena are induced by noise in time rather than by noise in space.

The general differential equation of the class of delayed dynamics with stochastic time is 
\begin{equation}
{dx(\bar{t}) \over d\bar{t}} = f(x(\bar{t}), x(\bar{t}-\tau)).
\end{equation}

Here, $x$ is the dynamical variable, and $f$ is 
the ``dynamical function'' governing the dynamics.  $\tau$ is the delay. The difference from the 
normal delayed dynamical equation appears in ``time'' $\bar{t}$, which contains stochastic 
characteristics.  We can define $\bar{t}$ in a variety of ways as well as the 
function $f$.
To avoid ambiguity and for simplicity, we focus on the following dynamical 
map system incorporating the basic ideas of the above general definition. 
\begin{eqnarray}
x_{n_{k+1}} & = & f(x_{n_k}, x_{{n_k}-\tau}),\nonumber\\
n_{k+1} & = & n_k + \xi_k
\end{eqnarray}
Here, $\xi_k$ is the stochastic variable which can take either $+1$ or $-1$
with certain probabilities. We associate ``time'' with an integral variable $n$. The dynamics progress by incrementing integer
$k$, and $n$ occasionally ``goes back a unit'' with the occurrence of $\xi=-1$.
Let the probability of $\xi_k =-1$ be $p$ for all $k$, and we set $n_{0}=0$. Then naturally, with $p=0$,
this map reduces to a normal delayed map with $n_k=k$. We update the variable $x_n$ with the larger $k$.
Hence, $x_n$ in the ``past''
could be ``re-written'' as $n$ decreases with the probability $p$. 

Qualitatively, we can make an analogy of this model with a tele--typewriter or a tape--recorder, which
occasionally moves back on a tape. A schematic view is shown in Figure 1A. The recording head writes on
the tape the values of $x$ at a step, and ``time'' is associated with  positions on the tape.
When there is no fluctuation ($p=0$), the head moves only in one direction on the tape and it records  values of $x$ for a normal delayed dynamics.
With probability $0 < p$, it moves back a unit of ``time'' to
overwrite the value of $x$. The question is how the recorded patterns of $x$ on the tape are affected 
as we change $p$.

The dynamical function is
chosen to be a negative feedback function (Figure 1B) and the concrete map model that we
will study is given as follows.
\begin{equation}
x_{n_{k+1}} = x_{n_k} + d\delta (-\alpha x_{n_k} - {2 \over {1+e^{-\beta x_{{n_k}-\tau}}}} +1),
\end{equation}
with $\alpha$, $\beta$ and $d\delta$ as parameters. With both $\alpha$ and $\beta$ positive and
no stochasticity in time,
this map has the origin as a stable fixed point
with no delay.  We consider the case of $\alpha =0.5$, $\beta =6$, and $d\delta=0.1$. By a linear stability analysis, the critical delay $\tau_c$, at which the stability of the fixed point is
lost, is given as $\tau_c \sim 6$. The larger delay gives an oscillatory dynamical path.
We have found, through computer simulations, that an interesting behavior
arises when delay is smaller than this critical delay. The tuned noise in the time
flow gives the system a tendency for oscillatory behavior. In other words,
adjusting the value of $p$ controlling $\xi$ induces an oscillatory dynamical path.
Some examples are shown in Figure 1C. With increasing probability for the time
flow to reverse, i.e., with $p$ increasing, we observe oscillatory behavior
both in the sample dynamical path as well as in the corresponding power spectrum.
However, when $p$ reaches beyond an optimal value, the oscillatory behavior begins to 
deteriorate. The change in the peak heights is shown in Figure 1D. 
This phenomenon resembles stochastic resonance. 
A resonance with delay and noise, called
``delayed stochastic resonance''\cite{ohira-sato}, has been proposed for an additive noise in
``space''. Analytical understanding 
of the mechanism  is yet to be explored for our model. However,
this mechanism of stochastic time flow is clearly of a different type and new. 

\begin{figure}
\includegraphics[width=.85\textwidth]{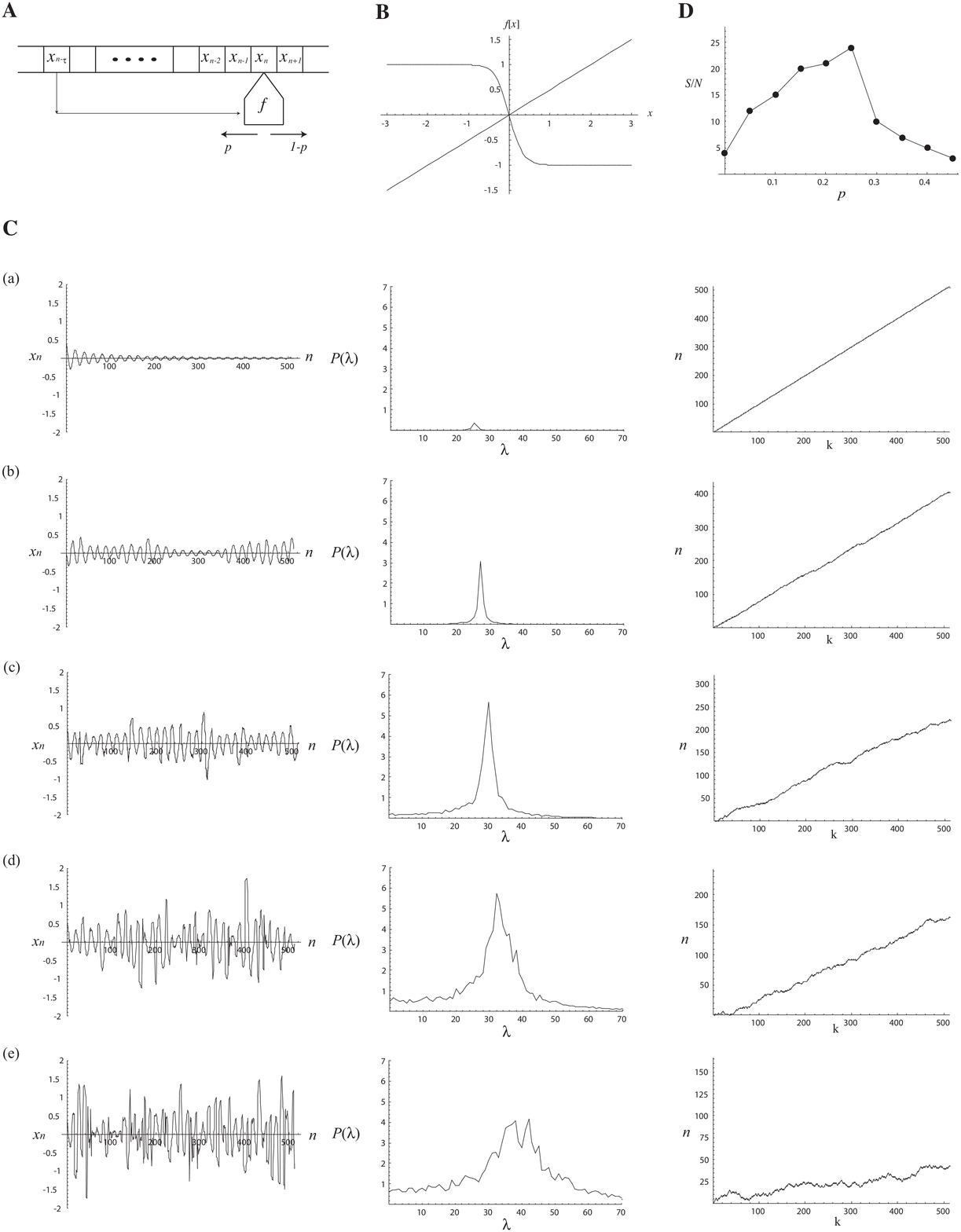}
\caption{
{\bf A}: A schematic view of the model.
{\bf B}: Dynamical functions $f(x)$ with parameters as examples of simulations presented in this paper. Negative feedback function with parameters $\beta = 6$. Straight line has 
slope of $\alpha = 0.5$.
{\bf C}: Dynamics (left) and power spectrum (middle) of delayed dynamical model with 
stochastic time flow (right). This is an example of the dynamics and associated power 
spectrum through the simulation of the model given in Eq. (3) with the probability 
$p$ of stochastic time flow varied.  The 
parameters are set as $\alpha = 0.5$, $\beta = 6$, $d\delta=0.1$, $\tau=5$ and the 
stochastic time flow parameter $p$ are set to (a) $p=0$, (b) $p=0.1$, (c) $p=0.25$, (d) 
$p=0.35$, and (e) $p=0.45$. We used the initial condition that $x_{n} = 0.5 (n\leq 0)$, and
 ${n_0}=0$.
The simulation is performed up to $k=5000$ steps. At that point the values of $x_n$ for $0 \leq n \leq L$ with
$L=512$ are recorded. The 50 averages are taken for the power spectrum on this recorded $x_n$. The unit of frequency $\lambda$ is set as ${1 \over L}$, 
and the power $P(\lambda)$ is in arbitrary units. 
{\bf D}: The signal to 
noise ratio ${S \over N}$ at the peak height as a function of the probability  
$p$ of stochastic time flow. The parameter settings are the same as in {\bf C}.}
\end{figure}

We would like to now discuss a couple of points with respect to our model.
First, we can view this model as a dynamical model with non-locality and fluctuation
on time axis. Both factors are familiar in ``space'', but not on time.
We may extend this model to include non-locality and fluctuations in the space variable 
$x$. Proceeding in this way, we have a picture of dynamical systems with 
non-locality and fluctuations on both the time and space axes. The analytical framework and 
tools for such a description need to be developed, along with a search for 
appropriate applications. 

Another 
way might be to extend the path integral formalism. 
The question of whether this extension bridges to quantum mechanics and/or leads to 
an alternative understanding of such properties as time-energy uncertainty 
relations also requires further investigation. Also, there is a theory of elementary
particles with a fluctuation of space--time, where the noise term is added to
the metric\cite{takano}. If we can connect our ideas here to such a theory remains to
be seen.

Finally, if these models do capture some aspects of reality, particularly with respect to
stochasticity in the time flow, this resonance 
may be used as an experimental indication for probing fluctuations or 
stochasticity in time. We have previously proposed ``delayed stochastic 
resonance''\cite{ohira-sato}, a resonance that results from the interplay of noise and delay. 
It was theoretically extended\cite{tsimring01}, and recently, the effect was 
experimentally observed in a solid-sate laser system with a feedback 
loop\cite{masoller}. It is left for the future to see if an analogous 
experimental test could be developed with respect to stochastic time.

\end{document}